\documentclass{IEEEcsmag}

\usepackage[colorlinks,urlcolor=blue,linkcolor=blue,citecolor=blue]{hyperref}
\usepackage{upmath}
\usepackage{tikz}
\usetikzlibrary{arrows.meta, shapes, shapes.geometric,arrows,fit,matrix,positioning,chains,tikzmark}
\usepackage{pgfplots}
\usepackage{smartdiagram}
\usepackage{listings}
\usepackage[braket, qm]{qcircuit}
\usepackage{graphicx}
\usepackage{multirow}
\jvol{XX}
\jnum{XX}
\paper{8}
\pubyear{2022}

\usepackage{xcolor}

\begin{document}


\title{Quantum Circuit Transformations with a Multi-Level Intermediate Representation Compiler}

\author{T.~Nguyen}
\affil{Oak Ridge National Laboratory}

\author{D.~Lyakh}
\affil{Oak Ridge National Laboratory}

\author{R.~C.~Pooser}
\affil{Oak Ridge National Laboratory}

\author{T.~S.~Humble}
\affil{Oak Ridge National Laboratory}

\author{T.~Proctor}
\affil{Sandia National Laboratories}

\author{M.~Sarovar}
\affil{Sandia National Laboratories}

\markboth{Department Head}{Paper title}

\begin{abstract}
Quantum computing promises remarkable approaches for processing information, but new tools are needed to compile program representations into the physical instructions required by a quantum computer. Here we present a novel adaptation of the multi-level intermediate representation (MLIR) integrated into a quantum compiler that may be used for checking program execution. We first present how MLIR enables quantum circuit transformations for efficient execution on quantum computing devices and then give an example of compiler transformations based on so-called mirror circuits. We demonstrate that mirror circuits inserted during compilation may test hardware performance by assessing quantum circuit accuracy on several superconducting and ion trap hardware platforms. Our results validate MLIR as an efficient and effective method for collecting hardware-dependent diagnostics through  automated transformations of quantum circuits.
\end{abstract}

\maketitle

\chapterinitial{Quantum computing} has emerged in recent years as a revolutionary approach to processing information using the principles of quantum mechanics.
The quantum computational model is known to offer remarkable advances in algorithmic speed up for a number of problems that challenge conventional models, and there are many efforts to demonstrate quantum computational advantage. 
This includes theoretical results such as the algorithmic speedup available for factoring integers as well as experimental results such as the recent experimental demonstrations of so-called `quantum supremacy' \cite{arute2019quantum}.
The success of these efforts depends on transforming algorithmic statements made within the model of quantum computing into instructions carried out by control hardware linked to quantum devices. Whereas a programmer routinely expresses quantum algorithms using a variety of high-level languages \cite{heim2020quantum}, the translation of these statements into executable instructions benefits greatly from the use of automated compilation tools. 
\par
Transforming quantum algorithms into physical instructions relies on compiling high-level languages into sequences of operations that act on the information stored in a quantum hardware register \cite{humble2019quantum}. There are a variety of technologies being developed for quantum registers, including superconducting transmon qubits, trapped ions or atoms, and atomically doped silicon. Each technology supports a unique set of physical instructions and imposes a unique set of constraints that limit how these instructions may be executed. Tools that automate the decisions and translation of algorithms into technology-specific representations are vital for ensuring efficient and accurate execution \cite{chong2017programming}.
\par
A natural and easy-to-understand representation of the logical instructions executed during a quantum algorithm is the quantum circuit. An example of a quantum circuit is shown in Fig.~\ref{fig:quantum_circuit_1} to highlight a series of gates acting on a pair of quantum register elements. Additional notation is defined below. This diagrammatic representation emphasizes the placement and assignment of instructions acting on individual registers, but it does not easily convey the temporal and spatial constraints imposed by a given technology.
In addition, transforming such circuit representations into alternative logically equivalent forms requires the use of known rewriting rules. The quantum circuit representation has therefore been augmented with a variety of intermediate representations, which have been developed to represent and transform these quantum instructions.
\par
Here we extend a novel adaptation of the multi-level intermediate representation (MLIR) for compiling quantum programs. MLIR was recently put forth by Nguyen and McCaskey as a powerful paradigm for expressing, parsing, and transforming quantum circuit representations at both the logical and physical levels \cite{mlir_quantum_dialect}. 
Using LLVM infrastructure, 
their approach uses MLIR to demonstrate efficient transformations of a variety of quantum circuits. 
\par 
We expand this use of MLIR to now implement transformations that create circuits for testing quantum circuit executions on noisy quantum computing hardware. These methods for verification and validation of quantum computers are based on a form of circuit transformation known as mirror circuits \cite{arXiv2109.00506}. We present details on the methodology and implementation of these transformation alongside experimental results obtained from compiling test circuits to several different quantum computing hardware targets. These methods reveal characteristic differences in technologies and demonstrate the utility of combining MLIR with quantum compilation. 
\section{Quantum Compiling}
\par 
The quantum circuit representation is a useful, human readable, and diagrammatic description of quantum computer programs. At the machine level, specific intermediate representations (IR), often not intended for human reading, and native to specific architectures, are used. Various IRs, MLIR included, can be derived from quantum assembly language (QASM), which itself is frequently compiled from higher level circuit representations. The IR thus represents an endpoint in a tool-chain or compiler/transpiler stack that reduces high-level user input, often expressed in a quantum circuit paradigm, to a machine-readable format. Many current quantum computer programming and compilation tool stacks
offer such an IR endpoint.
\par 
Within this compiler concept, users often think of quantum programs visually. Users may envision a program as a circuit directly, and subsequently take on the task of translating the circuit into a high level language such as qiskit, which is then compiled (and in many cases, transpiled) by the remainder of the qiskit toolchain. It is useful to outline in brief then, the concepts contained at this high level, before outlining their reduction to MLIR. 
\par
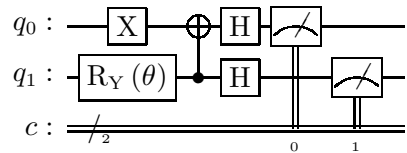
\begin{figure}
\scalebox{1.0}{
\Qcircuit @C=0.4em @R=0.2em @!R { \\
	 	\nghost{ {q}_{0} :  } & \lstick{ {q}_{0} :  } & \gate{\mathrm{X}} & \targ & \gate{\mathrm{H}} & \meter & \qw & \qw & \qw\\ 
	 	\nghost{ {q}_{1} :  } & \lstick{ {q}_{1} :  } & \gate{\mathrm{R_Y}\,(\theta)} & \ctrl{-1} & \gate{\mathrm{H}} & \qw & \meter & \qw & \qw\\ 
	 	\nghost{c:} & \lstick{c:} & \lstick{/_{_{2}}} \cw & \cw & \cw & \dstick{_{_{0}}} \cw \cwx[-2] & \dstick{_{_{1}}} \cw \cwx[-1] & \cw & \cw\\ 
\\ }}
\caption{An example of a simple quantum circuit diagram that illustrates a two-qubit register acted upon by a series of one- and two-qubit gates. The placement and assignment of gates to individual register elements is an important role for a quantum compiler. Notation is defined in the text.}
\label{fig:quantum_circuit_1}
\end{figure}

 
In Fig.~\ref{fig:quantum_circuit_1}, the \textit{qubit register} is denoted by the labels ${q_i}$, while the corresponding classical register, which is used to store the values obtained after quantum measurements, is denoted ${c}$. These two registers differ greatly in terms of the physics of their representation of information: a qubit exists in a continuum of possible states on the Bloch sphere, with states along the surface of the sphere representing superpositions of the states ``0'' and ``1'' (which are themselves represented by the poles). The classical register, on the other hand, consists of the traditional binary, or ``bit'' representation of information, in which each information carrier can exist in either the state ``0'' or ``1'', but not superpositions thereof (this also rules out concepts like entanglement from the classical register). 

Quantum circuits typically end with the measurement gate performed on a portion of the quantum register, and subsequent storage in the classical register for return to classical computer resources. The measurements report the resulting quantum computation that occurs throughout the circuit. Key elements of the quantum circuit include quantum gates, often acting on single and two-qubit subsets, although three or more qubit quantum gates are possible and can be  expressed in this notation. Figure~\ref{fig:quantum_circuit_1} shows several of the key gate components required for universal quantum computation: quantum ``bit flip'' ($X$) gates, Hadamard ($H$) gates, which are specific rotations along the Bloch sphere, rotations about specific axes of the sphere (see $R_y(\theta)$ in Fig.~\ref{fig:quantum_circuit_1}), and finally the controlled-not gate (CNOT, represented by the node, or control, connected to the crossed-circle, representing the target). These quantum logic gates are somewhat analagous to their classical logical counterparts with a key exception. The truth table of the CNOT gate, when applied to quantum superpositions, results in ``entanglement'', which means that the two qubits involved in the gate have become quantum-correlated. Measurements reveal these correlations, while subsequent quantum gates in the circuit may exploit these correlations to establish key advantages in quantum programs over their classical counterparts. The circuit diagram, and the resulting MLIR can fully capture this abstract concept within their logical notations.
\par 
\par 
\section{Quantum Circuit Transformations}
\par
\begin{figure}
\scalebox{0.7}{
\Qcircuit @C=0.3em @R=0.2em @!R { \\
	 	\nghost{ {q}_{0} :  } & \lstick{ {q}_{0} :  } & \gate{\mathrm{U_3}\,(\mathrm{\pi,0,-\pi})} & \targ & \gate{\mathrm{U_2}\,(\mathrm{0,\pi})} & \meter & \qw & \qw & \qw\\ 
	 	\nghost{ {q}_{1} :  } & \lstick{ {q}_{1} :  } & \gate{\mathrm{U_3}\,(\theta,\mathrm{0,0})} & \ctrl{-1} & \gate{\mathrm{U_2}\,(\mathrm{0,\pi})} & \qw & \meter & \qw & \qw\\ 
	 	\nghost{c:} & \lstick{c:} & \lstick{/_{_{2}}} \cw & \cw & \cw & \dstick{_{_{0}}} \cw \cwx[-2] & \dstick{_{_{1}}} \cw \cwx[-1] & \cw & \cw\\ 
\\ }}
\caption{An example of a circuit diagram illustrating the transformation of gates. Gates from the circuit in Figure~\ref{fig:quantum_circuit_1} is transformed into the native gate set of the accelerator backend. In this case, we use the IBM's \{$U_1$, $U_2$, $U_3$, $CX$\} gate set for demonstration purposes.}
\label{fig:quantum_circuit_2}
\end{figure}
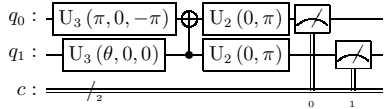

The expression of a given quantum algorithm in terms of quantum logic gates acting on a qubit register is not unique as the same composite unitary operation can be decomposed into a product of simpler unitary operations in many ways. Since different quantum devices provide different sets of native quantum gates implemented in physical hardware, the final form of a gate-based IR for a given quantum circuit must consist of only those gates which are natively implemented by the target quantum processing unit (see Fig. \ref{fig:quantum_circuit_2}). Additionally, the final gate-based IR must respect the physical connectivity between qubits, thus necessitating the insertion of the SWAP gates into the IR such that all 2-qubit quantum gates act on physically connected qubits (qubits that can directly interact and entangle with each other). 
\par 
Given physical hardware constraints, a typical quantum circuit IR transformation is aimed at reducing the depth of the circuit and/or the total number of elementary gates, especially single-qubit $T$ gates and multi-qubit entangling gates, which is a hard optimization problem with complexity growing rapidly with the circuit size \cite{npjQuantumInformation_4_23_2018, arXiv2106.11246}. Importantly, for some quantum algorithms such quantum circuit transformations result in a constant depth IR \cite{arXiv2108.03282, 2108.03283}. Reducing the depth of the quantum circuit typically increases the fidelity of its output on a noisy quantum computing device. However, different quantum gates have different fidelities or error rates, and accounting for this requires a more complex circuit transformation with an optimization functional based on individual gate fidelities \cite{arXiv2012.09835}. 
\par 
Traditionally, quantum circuit transformations include the following stages: circuit synthesis, postsynthesis optimization, and technology mapping \cite{ACMComputingSurveys_45_21_2013}. Such multi-stage quantum circuit transformations can benefit from gradual IR lowering introduced by MLIR. Below we describe an additional class of quantum circuit transformations aimed at validation of quantum processing units. This class of quantum circuit transformations, called mirror circuit transformations, is automated via our advanced quantum compiler technology based on MLIR \cite{arXiv2109.00506}.

\section{Mirror Circuit Transformations}
\par 
Mirror circuits are a method for estimating the error in a quantum processor's implementation of some circuit $C$ \cite{Proctor2020-ky}. Mirror circuits are designed to address the problem that the output of $C$ cannot be efficiently verified as correct in general. This is because a general quantum circuit $C$ cannot be efficiently simulated on a classical computer, so the output from running a quantum circuit $C$ on a quantum computer cannot be compared to what we would have got if $C$ had been run without any errors. Mirror circuits are circuits that are similar to $C$, but whereby each mirror circuit will always output the same easily predictable bit string if run without error. This makes it easy to quantify the error in the execution of these circuits, and from this we can approximate the error with which $C$ has or could be implemented.
\par 
A set of mirror circuits $\{M(C)\}$ is constructed from a ``base'' circuit, which is the circuit we want to quantify a particular quantum processor's performance on. The version of mirror circuits used in this work is introduced in Ref.~\cite{Proctor2020-ky} and it is defined for any circuit containing only Clifford gates and $Z(\theta)$ gates (arbitrary rotation around the $Z$ axis). This encompasses many widely used universal gate sets. A mirror circuit consists of a ``compute-then-uncompute'' structure alongside randomized elements that are added in to guarantee sensitivity to all errors that impact $C$ (some kinds of coherent errors can be echoed away by a compute-then-uncompute Loschmidt echo). Each mirror circuit $M(C)$ consists of (1) the circuit $C$, (2) a layer of random Pauli gates, (3) a ``quasi-inverse'' circuit, (4) a measurement of each qubit in the computational basis. The ``quasi inverse'' consists of $C$ in reverse with each gate replaced with its inverse except that the angles of the $Z(\theta)$ gates are updated to commute the central Pauli through the circuit. A complete description is given in Ref.~\cite{Proctor2020-ky}. 
\par 
Each mirror circuit $M(C)$ has a corresponding ``target'' bit string $s$ that it will always output if the circuit is run without errors. The theory in Ref.~\cite{Proctor2020-ky} shows that the probability of observing this bit string ($P$) is closely related to the fidelity with which the tested processor can implement $C$. In this work, we use a performance metric for $C$ computed from  $\epsilon = (1-\sqrt{P}) \approx  (1 - P)/2$. We square root $P$ to remove the contribution of errors in the quasi-inverse circuit to $P$'s deviation from 1, so $\epsilon$ is then an estimate for the error in $C$. In particular, we report the mean and minimum of $\epsilon$ over the ensemble of (random) mirror circuits constructed and executed alongside a given $C$.

\begin{figure}
\scalebox{0.54}{
\Qcircuit @C=0.2em @R=0.2em @!R { \\
	 	\nghost{ {q}_{0} :  } & \lstick{ {q}_{0} :  } & \gate{\mathrm{U_3}\,(\mathrm{-\pi,-\frac{\pi}{2},-\frac{\pi}{2}})} & \targ & \gate{\mathrm{U_3}\,(\mathrm{\frac{\pi}{2},0,0})} & \gate{\mathrm{U_3}\,(\mathrm{-\frac{\pi}{2},\pi,\pi})} & \targ & \gate{\mathrm{U_3}\,(\mathrm{-\pi,\frac{\pi}{2},\frac{\pi}{2}})} & \meter & \qw & \qw & \qw\\ 
	 	\nghost{ {q}_{1} :  } & \lstick{ {q}_{1} :  } & \gate{\mathrm{U_3}\,(\pi + \theta,\mathrm{0,0})} & \ctrl{-1} & \gate{\mathrm{U_3}\,(\mathrm{\frac{\pi}{2},0,0})} & \gate{\mathrm{U_3}\,(\mathrm{-\frac{\pi}{2},\pi,0})} & \ctrl{-1} & \gate{\mathrm{U_3}\,(-\theta,\mathrm{0,0})} & \qw & \meter & \qw & \qw\\ 
	 	\nghost{c:} & \lstick{c:} & \lstick{/_{_{2}}} \cw & \cw & \cw & \cw & \cw & \cw & \dstick{_{_{0}}} \cw \cwx[-2] & \dstick{_{_{1}}} \cw \cwx[-1] & \cw & \cw\\ 
\\ }}
\caption{Example of a mirror circuit transformation applied to the quantum circuit from Fig.~\ref{fig:quantum_circuit_2} showing a single realization of the Pauli-randomized mirror circuit. The transformed circuit has the same structure (double depth) as that of the original yet is expected to produce a deterministic output bit string, e.g.,`11', for validation purposes. }
\label{fig:quantum_circuit_3}
\end{figure}
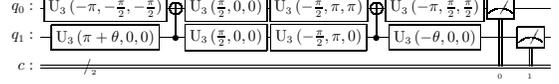

\section{Quantum Circuit Transformations using MLIR with QCOR}
\par
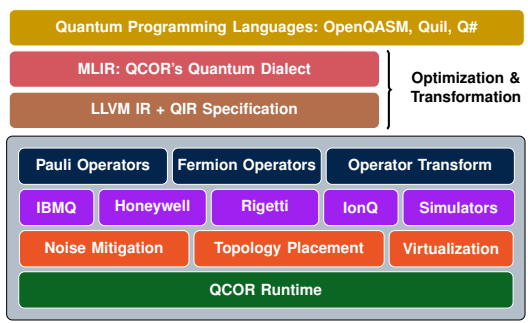
\begin{figure}[b]
\centerline{
\scalebox{0.55}{%
\definecolor{mybluei}{RGB}{3, 37, 76}
\definecolor{myblueii}{RGB}{73,121,193}
\definecolor{mygreen}{RGB}{11, 102, 35}
\definecolor{myred}{RGB}{160, 32, 240}
\definecolor{mlir}{RGB}{212, 87, 96}
\definecolor{mybrown}{RGB}{179, 110, 75}
\definecolor{dsl}{RGB}{201,151,0}
\definecolor{orange}{RGB}{235, 84, 36}
\pgfdeclarelayer{background}
\pgfsetlayers{background,main}
\begin{tikzpicture}[node distance=3pt,
blueb/.style={
  draw=white,
  fill=mybluei,
  rounded corners,
  text width=2.5cm,
  font={\sffamily\bfseries\color{white}},
  align=center,
  minimum height = 25pt,
  },
greenb/.style={blueb,fill=mygreen},
redb/.style={blueb,fill=myred},
orangeb/.style={blueb,fill=orange},
]
\node[blueb, text width=11.5cm+20pt, fill=dsl] (DSL) {Quantum Programming Languages: OpenQASM, Quil, Q\#};
\node[blueb, text width=8.0cm+20pt, fill=mlir, xshift = -1.75cm, below=of DSL] (MLIR) {MLIR: QCOR's Quantum Dialect};
\node[blueb, text width=8.0cm+20pt, fill=mybrown, below=of MLIR] (LLVM) {LLVM IR + QIR Specification};

\node[blueb, text width=3.0cm+20pt, text height=1.15cm, draw=none, fill=none, font=\sffamily\bfseries, right=of MLIR] (Opt) {Optimization \& Transformation};
\draw [decorate, decoration = {brace}, line width=0.5mm] ([xshift=5pt,yshift=0pt]MLIR.north east) --  ([xshift=5pt,yshift=0pt]LLVM.south east);
\node[blueb, below=of DSL, xshift=-4.2cm, yshift=-2.35cm, text width=3cm+10pt] (Pauli) {Pauli Operators};
\node[blueb, right=of Pauli, text width=3cm+10pt] (Fermion) {Fermion Operators};
\node[blueb, right=of Fermion, text width=3.95cm+10pt] (OpTransform) {Operator Transform};
\node[redb, text width=1.2cm+10pt, below= of Pauli, xshift=-25pt] (IBMQ) {IBMQ};
\node[redb, right=of IBMQ, text width=2cm+10pt] (Honeywell) {Honeywell};
\node[redb, right=of Honeywell, text width=2cm+10pt] (Rigetti) {Rigetti};
\node[redb, right=of Rigetti, text width=1.2cm+10pt] (IonQ) {IonQ};
\node[redb, right=of IonQ, text width=2.1cm+10pt] (Sim) {Simulators};
\node[orangeb, below=of IBMQ, xshift = 1.15cm, text width=3.5cm+10pt] (noisemitigation) {Noise Mitigation};
\node[orangeb, right=of noisemitigation, text width=4cm+10pt] (placement) {Topology Placement};
\node[orangeb, right=of placement, text width=2.4cm+10pt] (virtualization) {Virtualization};
\node[greenb,below left=2.1cm and -8.25cm of Fermion,text width=11cm+20pt] (QRT) {QCOR Runtime};
\begin{pgfonlayer}{background}
\draw[blueb,draw=black,fill=mybluei!30] 
  ([xshift=-8pt,yshift=8pt]Pauli.north west) rectangle 
  ([xshift=8pt,yshift=-8pt]QRT.south east);
\end{pgfonlayer}

\end{tikzpicture}}
}

\caption{The QCOR software stack transforms high-level language expressions into an intermediate representation that drives hardware execution. The QCOR runtime provides services such as quantum job submission to remote quantum providers, pre and post-processing of execution data, and application-level primitives for quantum computing.   }
\label{fig:qcor_stack}
\end{figure}

We implement the transformation required by the mirror circuit theory using the QCOR programming framework. As shown in Figure~\ref{fig:qcor_stack}, QCOR is a full-stack quantum compilation and execution framework based on the LLVM's MLIR infrastructure. Specifically,  quantum programs in various languages are parsed and transformed into a unified MLIR leveraging the quantum dialect that was developed previously~\cite{mlir_quantum_dialect}. QCOR provides a collection of \emph{passes}, which represent the basic infrastructure for transformation and optimization within the MLIR. These passes are specific to quantum computing and includes techniques for quantum gate simplification and substitution.
\par
In MLIR, the transformation of the IR tree is managed by a \texttt{PassManager} that configures and schedules a pipeline of passes. For example, QCOR defines various optimization levels, each of which enables a different set of circuit optimization passes to balance the compilation runtime and the optimality of the resulting quantum executable. Passes operate on \texttt{Operation} instances which are the main abstraction of MLIR. For instance, sub-routine (function) level transformations are associated with MLIR's \texttt{FuncOp} representing callable function objects. Similarly, the global transformation of the whole program is defined by \texttt{ModuleOp} passes, whereby \texttt{ModuleOp} is MLIR abstraction for top-level container operation. 
\par 
A pertinent pattern of QCOR transformation passes is the IR tree traversal via \emph{static single assignment} (SSA) \texttt{use-def} chain tracing and in-place replacement or removal of MLIR \texttt{Operation} nodes. By following the \texttt{use-def} chain of qubit lines, we can establish the \emph{directed acyclic graph} (DAG) representation of the compiled program amenable to quantum circuit transformation protocols, e.g., gate optimization or circuit mirroring. Once a pass identifies an IR sub-tree matching its target pattern, it will perform the corresponding transformation, such as canceling quantum gate operations or replacing a sequence of gates by an equivalent yet more optimal substitute.   
\par
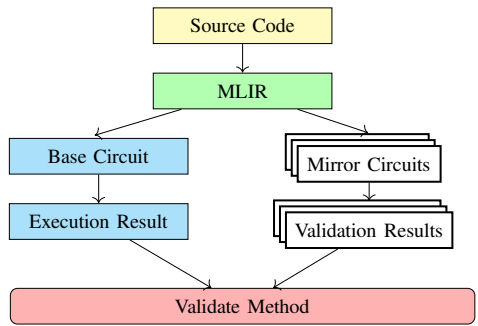
\begin{figure}[t!]
\scalebox{0.95}{%
\tikzset{
box/.style args = {#1/#2/#3}{rectangle,
        minimum width=#1, fill=#2!30, draw,
        text width =\pgfkeysvalueof{/pgf/minimum width}-2*\pgfkeysvalueof{/pgf/inner xsep},
        minimum height=#3, align=center,
        font=\footnotesize},
box/.default = 18mm/green/0.5cm,
}

\tikzset{%
  cascaded/.style = {%
    general shadow = {%
      shadow scale = 1,
      shadow xshift = -1ex,
      shadow yshift = 1ex,
      draw,
      thick,
      fill = white},
    general shadow = {%
      shadow scale = 1,
      shadow xshift = -.5ex,
      shadow yshift = .5ex,
      draw,
      thick,
      fill = white},
    fill = white, 
    draw,
    thick,
    font=\footnotesize,
    minimum width = 20mm,
    minimum height = 5mm}}
\tikzstyle{decision} = [diamond, aspect=2, draw, text badly centered, inner sep=3pt, font=\footnotesize]

\begin{tikzpicture}[node distance = 4mm and 6mm]

\node (src) [box=25mm/yellow/0.5cm] {Source Code};
\node (MLIR) [box=25mm/green/0.5cm, below=of src] {MLIR};
\node (Base) [box=25mm/cyan/0.5cm, below left= 4mm and -5mm of MLIR] {Base Circuit};
\node (Validate) [cascaded, below right= 5mm and -5mm of MLIR] {Mirror Circuits};
\node (BaseResult) [box=25mm/cyan/0.5cm, below = of Base] {Execution Result};
\node (MirrorResults) [cascaded, below  = of Validate] {Validation Results};
\node (decision) [box=65mm/red/0.5cm, rounded corners=1mm, below = 25mm of MLIR] {Validate Method};
\draw[->] (src) -- (MLIR);
\draw[->] (MLIR) -- ($(Base.south west)+(1.2, 0.55)$);
\draw[->] (MLIR) -- ($(Validate.south west)+(1.0, 0.7)$);
\draw[->] (Base) -- (BaseResult);
\draw[->] (Validate) -- ($(MirrorResults.north)+(0.0, 0.15)$);
\draw[->] (BaseResult) -- (decision);
\draw[->] (MirrorResults) -- (decision);


\end{tikzpicture}
}
\caption{MLIR transformation for V\&V of backend execution. When V\&V is enabled, \texttt{qcor} compiler will inject additional operations to the IR, and hence the resulting binary executable to run mirror circuits alongside the algorithmic circuit on the target QPU. The results of these mirror circuits are used to compute a fidelity estimate of the base circuit execution for V\&V purposes.}
\label{fig:qcor_mirror_diagram}
\end{figure}

 
Here, we focus attention on a specific type of IR transformation for \emph{verification and validation} (V\&V) based on the mirror circuit method outlined above. In digital computing technology, compilers often provide V\&V toolings, e.g., injecting extra debug symbols or instrumentation into the binary executable in certain build modes. These tools assist developers with error detection and localization, especially in complex codebases. In QCOR, we put forward a custom V\&V compiling mode for quantum programs leveraging the mirror circuit technique, as shown in Figure~\ref{fig:qcor_mirror_diagram}.
\par
The mirror circuit validation module consists of a compiler instrumentation transformation pass and a run-time library. The compile-time IR transformation modifies the MLIR tree to insert its instrumentation around the main quantum execution zone. It is worth noting that in an IR tree, there are a variety of node types declaring variables and functions, yet there should only be one so-called \emph{entry point} function defining the main program flow. Thus, in the validation build mode, QCOR instruments this entry point IR segment with LLVM function calls into the QCOR run-time library to facilitate the mirror circuit construction, execution, and post-processing. This results in a custom V\&V binary for the input quantum program whose execution involves both the original algorithmic quantum circuit and a set of corresponding Pauli-randomized mirror circuits (see Figure~\ref{fig:qcor_mirror_diagram}). Subsequently, the execution results, e.g.,  bit-string distribution, are annotated with validation data, such as the estimated success probability of the algorithmic circuit.

We use the example of the quantum circuit first presented in Figure~\ref{fig:quantum_circuit_1} using the diagrammatic representation and now presented in Figure~\ref{fig:qasm_code} as a snippet in a quantum program written in OpenQASM version 3.0. QCOR supports parsing of this high-level programming language into the MLIR.  
\begin{figure}[t]
\definecolor{codegreen}{rgb}{0,0.6,0}
\definecolor{codegray}{rgb}{0.5,0.5,0.5}
\definecolor{codepurple}{rgb}{0.58,0,0.82}
\definecolor{backcolour}{rgb}{0.95,0.95,0.92}
\definecolor{forestgreen}{rgb}{0,0.6,0}
\lstdefinelanguage{openqasm3}{
    keywordstyle=\color{magenta},
    numberstyle=\tiny\color{codegray},
    morestring=[b]",
    stringstyle=\color{codepurple},
    basicstyle=\ttfamily\footnotesize,
    breakatwhitespace=false,         
    breaklines=true,                 
    captionpos=b,                    
    keepspaces=true,                 
    numbers=left,                    
    numbersep=5pt,                  
    showspaces=false,                
    showstringspaces=false,
    showtabs=false,                  
    tabsize=2,
    comment = [l]{//},
    commentstyle = \color{forestgreen},
    keywords = {def, OPENQASM, include, reset, measure, barrier, x, h, ry, cx, qubit, bit}
}
\begin{lstlisting}[language=openqasm3]
// Filename: deuteron.qasm
OPENQASM 3;
include "stdgates.inc";

def ansatz(float[64]:theta) qubit[2]:q {
    bit c[2];
    x q[0];
    ry(theta) q[1];
    cx q[1], q[0];
    // Change basis to XX
    h q;
    c = measure q;
}

qubit q[2];
ansatz(0.297113) q;
\end{lstlisting}
\vspace{1mm}

\scalebox{0.55}{
\texttt{qcor \underline{\color{violet}-qpu\tikzmark{qpu} ibm:ibmq\_bogota} \underline{\color{blue}-shots\tikzmark{shots} 1024} {\color{red}-val\tikzmark{validate}idate} deuteron.qasm}
\begin{tikzpicture}[overlay,remember picture]
    \draw[arrows=->] 
    ( $ (pic cs:qpu) +(6pt,-2.5ex) $ ) -- 
    ( $ (pic cs:qpu) +(6pt,-0.8ex) $ );
    \node[anchor=north]
    at ( $ (pic cs:qpu) +(6pt,-2ex) $ )
    {Target backend};
    \draw[arrows=->] 
    ( $ (pic cs:shots) +(6pt,-2.5ex) $ ) -- 
    ( $ (pic cs:shots) +(6pt,-0.5ex) $ );
    \node[anchor=north]
    at ( $ (pic cs:shots) +(6pt,-2ex) $ )
    {Number of shots};
    \draw[arrows=->] 
    ( $ (pic cs:validate) +(6pt,4.0ex) $ ) -- 
    ( $ (pic cs:validate) +(6pt,1.8ex) $ );
    \node[anchor=north]
    at ( $ (pic cs:validate) +(6pt,+7ex) $ )
    {Enable validation IR transformation};
\end{tikzpicture}
}
\vspace{0.5mm}
\caption{OpenQASM 3.0 code (top) describing the quantum circuit presented in Figure~\ref{fig:quantum_circuit_1} and compilation command (bottom) using the \texttt{qcor} compiler. \texttt{qcor} compiles this source code into MLIR representation then LLVM IR and binary object code. During the MLIR-LLVM IR lowering phase, \texttt{qcor} applies a set of IR transformation passes, including the pass which incorporates mirror circuit validation into the circuit execution (\texttt{-validate} switch to turn on the mirror-circuit transformation pass). \texttt{-qpu} and \texttt{-shots} options specify the quantum accelerator backend and the number of shots to run, respectively.}
\label{fig:qasm_code}
\end{figure}
In this program, we define a simple quantum kernel, named \texttt{ansatz} (lines 5-13), to construct a parameterized quantum circuit designed to solve the Deuteron ground state energy with the \emph{variational quantum eigensolver} (VQE) algorithm \cite{dumitrescu2019deuteron}. In the main program flow, we allocate some qubits (line 15) then invoke the kernel with a randomly selected parameter value for demonstration purposes. 
\begin{table*}[!htbp]
\caption{Results of mirror circuit validation on different quantum hardware and emulator accelerators. $P'$ is the prediction for the circuit's success probability based on the success probability of mirror circuits. We run 32 mirror circuits and report the average and minimum values of $P'$. The probability distribution distance measures are computed against the theoretical distribution obtained by numerical simulation of the algorithmic circuit. Here, we present the Wasserstein (\textbf{W}) distance (also known as Earth mover's distance), the Energy (\textbf{E}) distance, the Kullback–Leibler (\textbf{KL}) and Jensen–Shannon (\textbf{JS}) distance for comparison.}
\centering
\begin{tabular}{|l | l | c | c | c | c | c | c |} 
 \hline
 \multirow{2}{2cm}{\textbf{Name}} & \multirow{2}{2cm}{\textbf{Descriptions}}  & \multirow{2}{1.0cm}{\textbf{mean($P'$)}} & \multirow{2}{1.0cm}{\textbf{min($P'$)}} & \multicolumn{4}{c|}{\textbf{Probability Distribution Distance}}\\
 \cline{5-8} & & & & \textbf{W} & \textbf{E} & \textbf{KL} & \textbf{JS} \\
 \hline
 HQS-LT-S1-SIM & Honeywell Quantum Systems, H1 Emulator &  0.992175 & 0.988212 & 0.009 & 0.071 & 0.035 & 0.018  \\   
 \hline
 HQS-LT-S2 & Honeywell Quantum Systems, H1 Hardware & 0.991576  & 0.980274 & 0.006 & 0.055 & 0.020 & 0.010  \\   
 \hline
 ibmq\_lima & IBM, Falcon r4T processor, 5 qubits & 0.951753 & 0.91714 & 0.016 & 0.089 & 0.046 & 0.023  \\
 \hline
 ibmq\_bogota & IBM, Falcon r4L processor, 5 qubits & 0.935946 & 0.913584 & 0.012 & 0.088 & 0.038 & 0.019  \\
 \hline
 IonQ-2019 & IonQ trapped ion processor (11 qubits) & 0.962767 & 0.855436 & 0.022 & 0.107 & 0.082 & 0.041   \\
 \hline
\end{tabular}
\label{table:validation_data}
\end{table*}
\par
Figure~\ref{fig:qasm_code} also demonstrates the use of the QCOR command-line executable \texttt{qcor} to compile the OpenQASM code. This command accepts a compiler switch, \texttt{-validate}, that enables the mirror circuit IR transformation by adding an MLIR pass that implements the instrumentation injection into the IR tree. In addition, as a retargetable compiler, QCOR targets a growing list of hardware backends via the \texttt{-qpu} flag. Passing the backend name in the format of \texttt{<provider name>:<device name>} instructs the compiler to output a program representation that can be executed on the given target, as described in more detail elsewhere \cite{mccaskey2020xacc}.
\par 
We test compilation and execution of the demonstration program on commercial quantum computing hardware from Honeywell Quantum Solutions, IBM, and IonQ as well as a numerical simulator. We test on the IBM \texttt{ibmq\_lima} and \texttt{ibmq\_bogota} devices, which are part of the second-generation Falcon family of transmon architectures. Public information from IBM
shows \texttt{ibmq\_bogota} and \texttt{ibmq\_lima} have a quantum volume of 32 and 8, respectively, with the former expected to be more performant. For trapped-ion devices, we test IonQ’s 11-qubit hardware developed in 2019 and the Honeywell Quantum Solutions 12-qubit H1 hardware labeled \texttt{HQS-LT-S2}. We also test the Honeywell Quantum Solutions emulator \texttt{HQS-LT-S1-SIM}, which is intended to provide a high-fidelity model for a closely related H1 device labeled \texttt{HQS-LT-S1}.     
\par 
In these tests, we configure the program to use 32 Pauli-randomized mirror circuits with each circuit sampled by 1024 measurement shots. This suffices for the small test circuit used here, but larger quantum circuits are expected to require additional randomized trials. The average and minimum success probability are recorded as $P' = \sqrt{P} = 1 - \epsilon$ with $\epsilon$ the error in the circuit.  The results of executing the code on all of these devices is shown in Fig.~\ref{fig:qasm_code}, and the mirror circuit validation mode on various backends are summarized in Table~\ref{table:validation_data}.
%
\par 
Our tests show that the results for success probability prediction $P'$ correlate with four commonly-used measures of statistical similarity. The latter compare experimental outcomes against the idealized results and, as shown in  Table~\ref{table:validation_data}, measures for i) the Wasserstein distance, ii) the energy distance, iii) Kullback–Leibler divergence, and iv) the Jensen–Shannon distance correlate with the success probability prediction. It is notable that these varying distance metrics weigh divergence between the ideal and actual results differently and, for the current data set, the measures all show the same trends.  
\par 
The in-situ validation information provided by mirror circuits is valuable for evaluating programs on near-term quantum devices. This information helps account for performance variations that arise between qubits on the same chip or from drift during calibration cycles. For example, the \texttt{ibmq\_bogota} device is reported to have a larger quantum volume than \texttt{ibmq\_lima}, but the latter demonstrated similar success probability $min(P')$ for the mirror-circuit validation runs. This is corroborated by similarities in the energy distance metric. However, many different factors may contribute to changes in the distance measures, including performance drift between execution of the base and mirror circuits and a proportional difference in terms of gate noise versus readout error between devices. The in-lined testing provided by the transformed circuits enable an efficient assessment of such differences. Our current analysis method for the mirror circuit data is known to be more accurate when gate errors dominate the overall circuit error, and future improvements will help quantify the impact of readout error.

\section{Conclusions}
\par 
We have presented the first example of using MLIR for automatic generation of quantum circuit transformations that allow for collection of additional information for verification and validation of quantum circuit execution. This additional information is collected via coordinated execution of additional mirror circuits that allow one to estimate the fidelity of the original application circuit's execution on a given hardware platform. Our implementation with the QCOR compiler generates the required hardware instructions to run the original application circuit as well as the mirror circuits upon activation by a compilation flag. We have validated our implementation on several available quantum devices and discovered notable trends with other measures of similarity that corroborate our results. 
\par
We anticipate that the MLIR, and its implementation, will prove useful for other tasks required by the verification, validation and debugging of quantum computers. These tasks require hardware-derived information about the quality of program execution that will become increasingly important as classical simulations of rapidly scaling quantum computers become intractable in the near future \cite{arute2019quantum, Huang_2020}. For example, recently introduced methods for using mirror circuits to provide debugging information about quantum circuits through local circuit inversions \cite{debugging} would natural map to the automated workflow enabled by using MLIR for circuit transformations. Our demonstrations of MLIR within in QCOR, and the accessibility of these transformation at compile, offer a robust and convenient methods for quantum programming. In summary, we advocate for the view that automated circuit transformations and implementation of debugging and verification tasks in a manner that requires little user intervention are promising applications of advanced quantum compilation techniques.


\section*{Acknowledgement}
This work was supported by the U.S.~Department of Energy (DOE) Office of Science Advanced Scientific Computing Research program office Accelerated Research for Quantum Computing (ARQC) program. Support for implementation on the IonQ hardware platforms was provided by the Quantum Testbed Pathfinder program (ERKJ332). This research used resources of the Oak Ridge Leadership Computing Facility, which is a DOE Office of Science User Facility supported under Contract DE-AC05-00OR22725.
This manuscript has been authored by UT-Battelle, LLC under Contract No. DE-AC05-00OR22725 and Lawrence Berkeley National Laboratory under Contract No. DE-AC02-05CH11231 with the U.S. Department of Energy. The United States Government retains and the publisher, by accepting the article for publication, acknowledges that the United States Government retains a non-exclusive, paid-up, irrevocable, world-wide license to publish or reproduce the published form of this manuscript, or allow others to do so, for United States Government purposes. The Department of Energy will provide public access to these results of federally sponsored research in accordance with the DOE Public Access Plan. (http://energy.gov/downloads/doe-public-access-plan).

\bibliographystyle{iopart-num}
\bibliography{main}
\begin{IEEEbiography}{Thien Nguyen}{\,} is a research scientist with Oak Ridge National Laboratory, USA. His research interests include quantum programming and simulation of quantum systems. Nguyen received a Ph.D. degree from the Australian National University, Canberra, Australia. He is a
member of IEEE. Contact him at nguyentm@ornl.gov.
\end{IEEEbiography}

\begin{IEEEbiography}{Dmitry Lyakh}{\,}is a research scientist at the Oak Ridge Leadership Computing Facility, Oak Ridge National Laboratory, USA. His research interests include hybrid quantum/classical high-performance computing frameworks and scalable classical simulations of quantum processing units at both the circuit and pulse levels. Lyakh received a Ph.D. degree in quantum chemistry from V.N.Karazin Kharkiv National University, Kharkiv, Ukraine. Contact him at liakhdi@ornl.gov.
\end{IEEEbiography}

\begin{IEEEbiography}{Raphael C.~Pooser}{\,}leads the Quantum Computing and Sensing group and is a distinguished research scientist at Oak Ridge National Laboratory. His research focuses on determining the performance limits of near-term quantum computers as well as building practical quantum devices. He received a PhD in quantum optics from University of Virginia. contact him at pooserrc@ornl.gov.
\end{IEEEbiography}

\begin{IEEEbiography}{Travis S.~Humble}{\,} is director at the Department of Energy’s Quantum Science Center, a distinguished scientist at Oak Ridge National Laboratory, and director of the lab’s Quantum Computing Institute. His research focuses on  developing quantum technologies and infrastructure for scientific discovery through quantum computing. He received a PhD in theoretical chemistry from University of Oregon. He is a senior member of IEEE. Contact him at humblets@ornl.gov.
\end{IEEEbiography}

\begin{IEEEbiography}{Timothy Proctor}{\,}is a research scientist at Sandia National Laboratories. His current research is focused on developing scalable and robust methods for benchmarking and characterizing quantum computers. He received his PhD in physics from the University of Leeds, UK. Contact him at tjproct@sandia.gov
\end{IEEEbiography}

\begin{IEEEbiography}{Mohan Sarovar}{\,}is a Quantum Information Scientist at Sandia National Laboratories, Livermore, CA, USA. He has broad expertise in quantum information science and technology and his current research interests are focused on developing quantum algorithms for science applications, near-term applications of quantum computing and simulation, and developing new techniques for characterizing quantum computers. He received a Ph.D. in theoretical physics from the University of Queensland, Australia. Contact him at mnsarov@sandia.gov.
\end{IEEEbiography}

\end{document}